\documentclass[twocolumn,amsmath,amssymb,10pt,superscriptaddress,a4paper,letterpaper,fleqn]{revtex4-1}
\usepackage{amssymb}
\usepackage{epsfig}
\usepackage{graphicx}
\usepackage{dcolumn}
\usepackage{array}
\usepackage{bm}
\usepackage{fancyheadings}
\usepackage{longtable}
\usepackage{multirow}
\usepackage{float}
\usepackage{afterpage}  
\usepackage{color}

\setlongtables
\usepackage[breaklinks=true,linkbordercolor={1 1 1}]{hyperref}

\begin{document}
\setcounter{page}{1}

\title{
\qquad \\ \qquad \\ \qquad \\  \qquad \\  \qquad \\ \qquad \\ 
MCNP6 Simulation of Quasi-Monoenergetic
$^7$Li(p,n) Neutron Sources below 150 MeV
}

\author{Stepan G. Mashnik}
\email[Corresponding author, electronic address:\\ ]{mashnik@lanl.gov}

\author{Jeffrey S. Bull}
\affiliation{Los Alamos National Laboratory, Los Alamos, NM 87545, USA }

\date{\today}

\begin{abstract}
The applicability of MCNP6 to simulate 
quasi-monoenergetic neutron sources from interactions of
proton beams with energies below 150 MeV on thick $^7$Li targets
have been studied. Neutron
spectra at zero degrees from a 2-mm $^7$Li layer
backed by a 12-mm carbon beam stopper in an Al flange bombarded
with protons of 20, 25, 30, 35, and 40 MeV have been calculated
with MCNP6 using the recent Los Alamos data library
as well as using the Bertini+Dresner and CEM03.03 event generators.
A comparison with the experimental neutron spectra shows that the event
generators do not do well in
describing such reactions, while 
MCNP6 using the LANL data library
simulates production of neutrons from p + $^7$Li 
in good agreement with the  measured data.
\end{abstract}
\maketitle


\lhead{ND 2013 Article $\dots$}
\chead{NUCLEAR DATA SHEETS}
\rhead{A. Author1 \textit{et al.}}
\lfoot{}
\rfoot{}
\renewcommand{\footrulewidth}{0.4pt}

\section{ INTRODUCTION}
A real monoenergetic neutron source in the MeV energy region is
not feasible, therefore ``quasi-monoenergetic'' neutron
sources are used in different applications. Among such sources, the ones
based on the $^7$Li(p,n) reaction are widespread.
Although this reaction was measured many times by different
authors at several laboratories
(see reviews on available data in Refs. \cite{leda,ProkofievND2001,IAEAMemo}),
the measured data do not cover continuously the whole incident proton
energy region and the whole kinematics region of secondary neutrons.
Therefore evaluated nuclear data libraries are needed for the $^7$Li(p,n) 
reaction, to be used in simulations with transport codes
of the ``quasi-monoenergetic'' neutron
production from thick lithium targets in every particular application. 

In Refs.\cite{leda,ProkofievND2001}, an ENDF-formated data library for
incident protons with energies up to 150 MeV was developed. 
In those papers, the
important $^7$Li(p,n$_0$) and $^7$Li(p,n$_1$) reactions were evaluated
from the experimental data, with their angular
distributions represented using Lengendre polynomial
expansions. The decay of the remaining reaction flux
was estimated from GNASH nuclear model calculations.
This leads to the emission of lower-energy neutrons and
other charged particles and gamma-rays from preequilibrium
and compound nucleus decay processes.
Examples of 
the use of these data in representative applications by a
radiation transport simulation with the code MCNPX \cite{MCNPX}
are also presented in Refs. \cite{leda,ProkofievND2001}. 

More recently, our data library was used successfully by 
Simakov et al.
in MCNPX simulations
to study the activation cross sections on
Bi, Au, Co, and Nb targets bombarded with quasi-monoenergetic neutrons
produced from the p+$^7$Li reaction at the NPI/\v{R}e\v{z} cyclotron
facility \cite{Simakov2011}. 
However, we never tested ourselves the p+$^7$Li data library 
with the latest Los Alamos
Monte Carlo transport code MCNP6 \cite{MCNP6}. 
To fill this gap, here we test the applicability of MCNP6 \cite{MCNP6}
to simulate  quasi-monoenergetic neutron
sources from interactions of
proton beams with energies below 150 MeV on thick $^7$Li targets.
We used in our study the neutron spectra measured recently by Uwamino et al.
from a 2 mm thick $^7$Li target bombarded with protons of 20, 25, 30, 35,
and 40 MeV \cite{Uwamino1997} and took advantages of an MCNPX input file
kindly sent to us by Dr. Simakov to simulate the geometry of our problem
(see Fig. 1), that we modified later for our MCNP6 needs.

\begin{figure}[!h]
\vspace*{2mm}
\includegraphics[width=0.9\columnwidth]{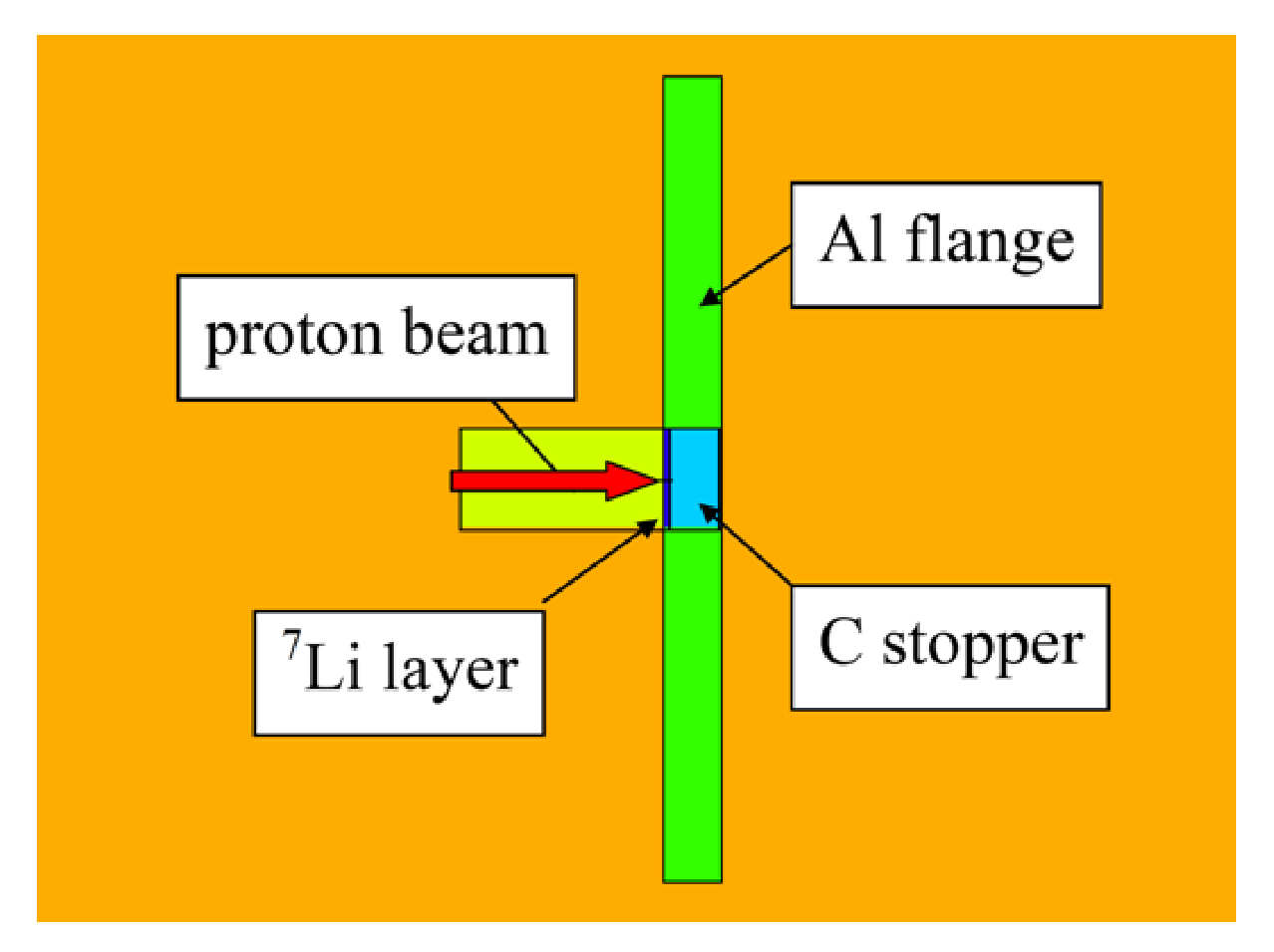}
\caption{Geometry model of the target and incident proton beam used
in our MCNP6 simulations. The plot was kindly sent to us
by Dr. Stanislav Simakov.
}
\label{fig1}
\end{figure}

\vspace*{-1mm}

\begin{figure*}[!htb]
\includegraphics[width=1.0\textwidth]{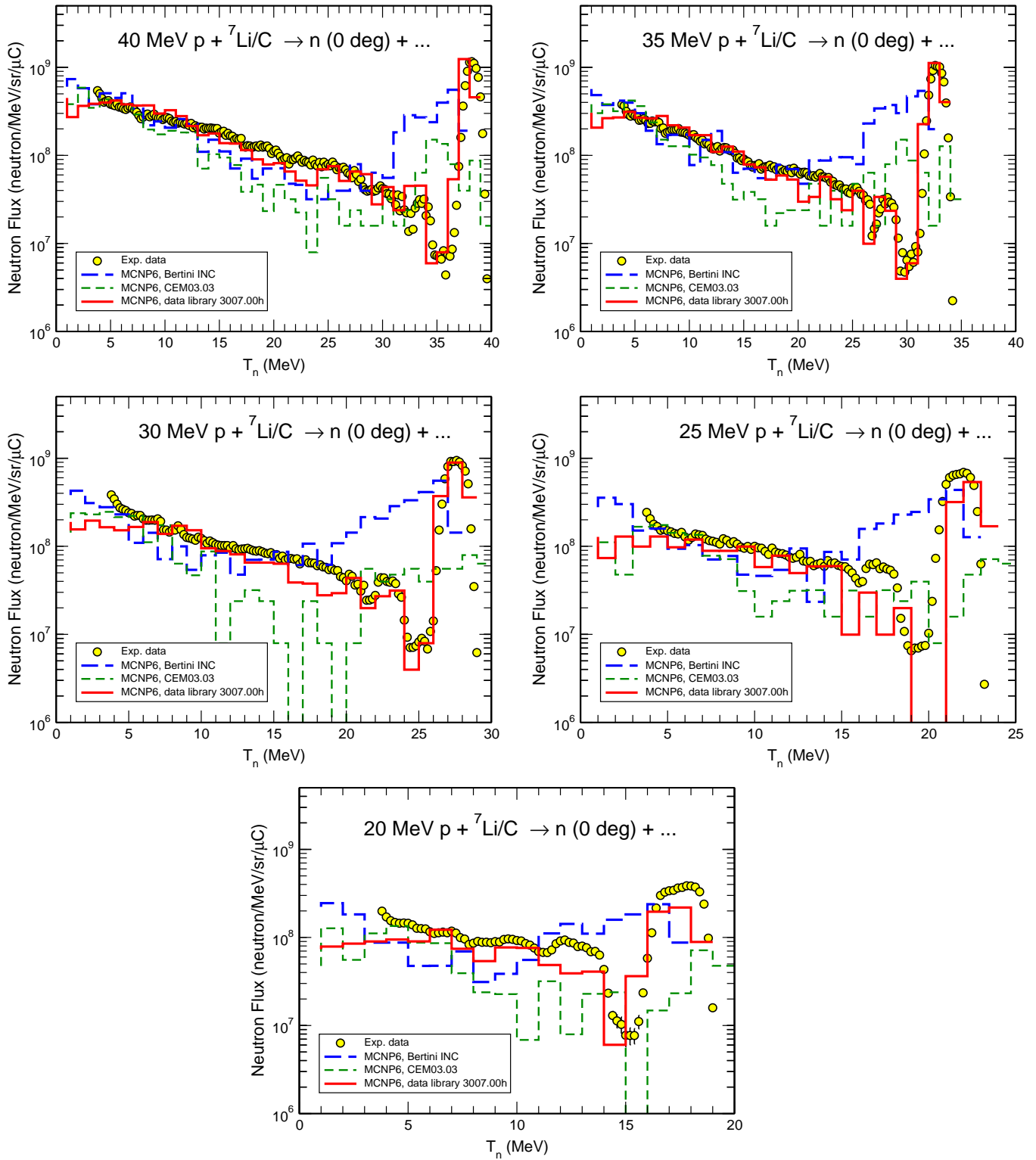}
\caption{Comparison of the MCNP6 results (histograms)
for neutron spectra from the thick $^7$Li target
(see Fig. 1 for details) bombarded with proton beams of
40, 35, 30, 25, and 20 MeV with the experimental data
by Uwamino et al. \cite{Uwamino1997} (solid circles). 
MCNP6 spectra calculated
using the p+$^7$Li data library \cite{leda} and, for comparison,
with the Bertini+Dresner \cite{Bertini,Dresner} and CEM03.03 \cite{Trieste08}
event generators 
are shown with solid red, blue dashed, and green
dashed histograms, as indicated in legends.
}
\label{fig3}
\end{figure*}

\clearpage

\section{Results}

As a step of MCNP6 Verification and Validation (V\&V), we have 
simulated the neutron spectra from our target consisting of a 2 mm thick $^7$Li
foil, 8 mm thick C beam stopper, and Al holder 
for the Li foils (see \cite{Uwamino1997} and Fig. 1 for details)
running MCNP6 both in a sequential mode and in parallel, with MPI,
while using our p+$^7$Li data library.
As expected, all results obtained with MCNP6
run with MPI coincide with the ones calculated in a sequential mode.
As additional steps of MCNP6 V\&V,
we have calculated all the studied neutron spectra
also using the Bertini+Dresner \cite{Bertini,Dresner} and CEM03.03 
\cite{Trieste08} event generators of MCNP6. All our results
are presented in Fig. 2. 
Just as expected for such low energies of incident protons,
we can see that neither the
Bertini+Dresner \cite{Bertini,Dresner} nor the newer CEM03.03 
\cite{Trieste08} event generators can reproduce satisfactorily
the measured neutron spectra from our $^7$Li target: The higher
the incident energy the better the agreement, but the quasi-monoenergetic
neutron peak is not reproduced well by the models
even at 40 MeV. At the same time, all the MCNP6 neutron spectra
calculated using the  p+$^7$Li data library agree well with the
measured data, including in the region of the quasi-monoenergetic
neutron peak of interest to our study, for all tested  
bombarding energies of the proton beam.

\vspace*{5mm}
\section{ CONCLUSIONS}
We have tested 
the applicability of MCNP6 to simulate 
quasi-monoenergetic neutron sources from interactions of
proton beams with energies below 150 MeV on thick $^7$Li targets. 
Neutron spectra at zero degrees from a 2-mm $^7$Li layer
backed by a 12-mm carbon beam stopper in an Al flange bombarded
with protons of 20, 25, 30, 35, and 40 MeV have been calculated
with MCNP6 using the recent Los Alamos data library
as well as using the Bertini+Dresner and CEM03.03 event generators.
Our results show that the models
(event generators) do not do well in
describing such reactions, while 
MCNP6 using the LANL data library
simulates production of neutrons from p + $^7$Li 
in good agreement with the measured data,
including the region of the quasi-monoenergetic
neutron peak of interest to our work.

We thank Drs. Stanislav Simakov and Naohiko Otsuka for useful discussions and
help on our study. 
This work was carried out under the auspices of the National Nuclear 
Security Administration of the U.S. Department of Energy at Los Alamos 
National Laboratory under Contract No. DE-AC52-06NA25396.

\end{document}